\documentclass[superscriptaddress,aps,prl,showkeys,showpacs,twocolumn,10pt]{revtex4}
\usepackage[OT1,T1]{fontenc}
\usepackage{graphicx}
\usepackage{rotating}
\begin{document}

\title{ABC Effect in Basic Double-Pionic Fusion --- Observation of
a new resonance?}
\date{\today}

\newcommand*{\IKPUU}{
Department of Physics and Astronomy, Uppsala University, 
75120 Uppsala, Sweden}
\newcommand*{\Erl}{Physikalisches Institut, Friedrich--Alexander--
Universit\"at Erlangen--N\"urnberg, 
91058 Erlangen, Germany}
\newcommand*{\ASWarsN}{Department of Nuclear Reactions, The Andrzej Soltan 
 Institute for Nuclear Studies, 
00-681 Warsaw, Poland}
\newcommand*{\IKPJ}{Institut f\"ur Kernphysik, Forschungszentrum J\"ulich, 
52425 J\"ulich, Germany}
\newcommand*{\ITEP}{Institute for Theoretical and Experimental Physics, 
State Scientific Center of the Russian Federation, 
117218 Moscow, Russia}
\newcommand*{\PITue}{Physikalisches Institut, Eberhard--Karls--Universit\"at
  T\"ubingen,  
72076 T\"ubingen, Germany}
\newcommand*{\MS}{Institut f\"ur Kernphysik, Westf\"alische  Wilhelms--
Universit\"at M\"unster, 
48149 M\"unster,  Germany}
\newcommand*{\ASWarsH}{High Energy Physics Department, The Andrzej Soltan 
 Institute for Nuclear Studies, 
00-681 Warsaw, Poland}
\newcommand*{\IITB}{Department of Physics, Indian Institute of Technology 
 Bombay, Powai, Mumbai--400076
, Maharashtra, India}
\newcommand*{\HISKP}{Helmholtz--Institut f\"ur Strahlen-- und Kernphysik, 
 Rheinische Friedrich--Wilhelms--
Universit\"at Bonn, 
53115  Bonn, Germany}
\newcommand*{\JCHP}{J\"ulich Center for Hadron Physics, Forschungszentrum 
 J\"ulich, 
52425 J\"ulich, Germany}
\newcommand*{\Bochum}{Institut f\"ur Experimentalphysik I, Ruhr--Universit\"at 
 Bochum, 
44780 Bochum, Germany}
\newcommand*{\IPJ}{Institute of Physics, Jagiellonian University, 
30-059 Krak\'{o}w, Poland}
\newcommand*{\TSL}{The Svedberg Laboratory, Uppsala University, 
75121 Uppsala, Sweden}
\newcommand*{\ZELJ}{Zentralinstitut f\"ur Elektronik, Forschungszentrum 
 J\"ulich, 52425 
J\"ulich, Germany}
\newcommand*{\Giess}{II.\ Physikalisches Institut, 
 Justus--Liebig--Universit\"at Gie{\ss}en,
35392 Giessen, Germany}
\newcommand*{\HepGat}{High Energy Physics Division, Petersburg Nuclear Physics 
 Institute, 
188300 Gatchina, Russia}
\newcommand*{\RWTH}{III.\ Physikalisches Institut~B, Physikzentrum, RWTH 
 Aachen, 52056 
Aachen, Germany}
\newcommand*{\IAS}{Institute for Advanced Simulation, Forschungszentrum 
 J\"ulich, 52425 
J\"ulich, Germany}
\newcommand*{\Katow}{August Che{\l}kowski Institute of Physics, University of 
 Silesia, 
40-007 Katowice, Poland}
\newcommand*{\IFJ}{The Henryk Niewodnicza{\'n}ski Institute of Nuclear 
 Physics, Polish Academy of Sciences, 
31-342 Krak\'{o}w, Poland}
\newcommand*{\Bethe}{Bethe Center for Theoretical Physics, Rheinische 
 Friedrich--Wilhelms--Universit\"at Bonn, 53115 
Bonn, Germany}
\newcommand*{\HiJINR}{Veksler and Baldin Laboratory of High Energiy Physics, 
 Joint Institute for Nuclear Physics, 
141980 Dubna, Russia}
\newcommand*{\IITI}{Department of Physics, Indian Institute of Technology 
 Indore, 
Indore--452017, 
Madhya Pradesh, India}
\newcommand*{\NuJINR}{Dzhelepov Laboratory of Nuclear Problems, Joint 
 Institute for Nuclear Physics, 
141980 Dubna, Russia}
\newcommand*{\ASLodz}{Department of Cosmic Ray Physics, The Andrzej Soltan 
 Institute for Nuclear Studies, 
90-950 Lodz, Poland}
\newcommand*{\INFN}{INFN, Laboratori Nazionali di Frascati,
00044 Frascati (Roma), Italy}
\newcommand*{\Wup}{Fachbereich Physik, Bergische Universit\"at Wuppertal, 
42119 Wuppertal, Germany}
\newcommand*{\FaJ}{Fachstrategie, Forschungszentrum J\"ulich, 52425 
J\"ulich,  Germany}

\author{P.~Adlarson}    \affiliation{\IKPUU}
\author{C.~Adolph}      \affiliation{\Erl}
\author{W.~Augustyniak} \affiliation{\ASWarsN}
\author{V.~Baru}        \affiliation{\IKPJ}\affiliation{\ITEP}
\author{M.~Bashkanov}   \affiliation{\PITue}
\author{F.S.~Bergmann}  \affiliation{\MS}
\author{M.~Ber{\l}owski}\affiliation{\ASWarsH}
\author{H.~Bhatt}       \affiliation{\IITB}
\author{K.--T.~Brinkmann}\affiliation{\HISKP}
\author{M.~B\"uscher}   \affiliation{\IKPJ}\affiliation{\JCHP}
% ...

\collaboration{WASA-at-COSY Collaboration}\noaffiliation






\begin{abstract}
We report on a high-statistics measurement of the
basic double pionic fusion reaction $pn \to d\pi^0\pi^0$ over the energy
region 
of the so-called ABC effect, a pronounced low-mass enhancement in the
$\pi\pi$-invariant mass spectrum. The measurements were performed with
the WASA detector setup at COSY. The data reveal the ABC effect  
to be associated with a Lorentzian shaped energy dependence in the integral
cross section. The observables are consistent with a resonance with $I(J^P)
=0(3^+)$ in both $pn$ and $\Delta\Delta$ systems. Necessary further tests of
the resonance interpretation are discussed.
\end{abstract}


\maketitle
The nature of the ABC effect, which denotes a pronounced low-mass enhancement in
the $\pi\pi$-invariant mass spectrum of double-pionic fusion reactions, has
been a puzzle all the time since its first observation 50 years ago by
Abashian, Booth and Crowe \cite{abc}. 
Follow-up experiments
revealed this
effect to be correlated with the production of an isoscalar pion pair. 
Previous interpretations were based on the mutual excitation of the two
colliding nucleons into the $\Delta(1232)$ resonance by meson exchange
($t$-channel $\Delta\Delta$ process), which leads to both a low-mass and a
high-mass enhancement in isoscalar $M_{\pi\pi}$ spectra
\cite{ris,barn,mos,alv}, in line with the 
 momentum distributions from inclusive measurements.

For the  reactions $pn \to d \pi^0\pi^0$ \cite{MB}, $pd \to
^3$He$\pi\pi$ \cite{he3} and $dd \to ^4$He$\pi\pi$ \cite{SK} the first
exclusive double-pionic fusion measurements of solid statistics carried out at
CELSIUS/WASA  firmly established the presence of 
a pronounced low $\pi\pi$--mass enhancement.
In addition, the first reaction showed a peculiar resonance-like behavior in
the energy dependence of the 
total cross section~ \cite{MB} --- indicative of a resonance in $pn$ and
$\Delta\Delta$ systems at a mass roughly 90 MeV below $2m_{\Delta}$ and a
width much smaller than expected from the conventional $t$-channel
$\Delta\Delta$ excitation process. 

On the other hand, in the {\it isovector} reactions $pp \to NN\pi\pi$ no
evidence for a significant ABC effect is found \cite{fk,TS,ts} -- including
the  limiting case of a fusion to a quasi-bound $pp$ pair in an $s$--wave  
\cite{ts,dym}. In all these cases the observables are well accounted for
by Roper and $t$-channel $\Delta\Delta$ processes.


In order to investigate this issue in a comprehensive way we measured the 
basic {\it isoscalar} double-pionic fusion process $pn \to d\pi^0\pi^0$
exclusively over the almost complete phase space with orders of magnitude
higher statistics than obtained previously \cite{MB}. 
The experiment was carried out with the WASA detector
setup \cite{wasa} at COSY 
via the reaction $pd \to d\pi^0\pi^0 + p_{spectator}$ using proton
beam energies of $T_p$ = 1.0, 1.2 and 1.4 GeV. Due to Fermi motion of the
nucleons in the target deuteron the quasifree reaction proceeds via a range of
effective collision energies with overlapping regions 2.22 GeV $\leq
\sqrt{s} \leq$ 2.36 GeV (triangles in Fig. 2), 2.33 GeV $\leq \sqrt{s} \leq$
2.44 GeV (dots) and 2.42 GeV $\leq \sqrt{s} \leq$ 2.56 GeV (squares).
The emerging deuterons were registered in the forward detector of WASA and
identified by the $\Delta$E-E technique. The photons from the $\pi^0$ decay were
detected in the central detector.

That way four-momenta were measured for all
particles of an event --- except for the very low-energetic spectator proton,
which did not reach any detector element. Since the reaction was 
measured kinematically overdetermined, the spectator momentum could be 
reconstructed and kinematic fits with 3 overconstraints performed for
each event.    
From the complete kinematic information available also the
relevant total energy in the pn system could be reconstructed for each event
individually.

The absolute normalisation of the data has been obtained by a
relative normalisation to quasifree $\eta$, $\pi^+\pi^0$ and $\pi\pi\pi$ 
production 
measured simultaneously with the same trigger.

\begin{figure} 
\centering
\includegraphics[width=0.99\columnwidth]{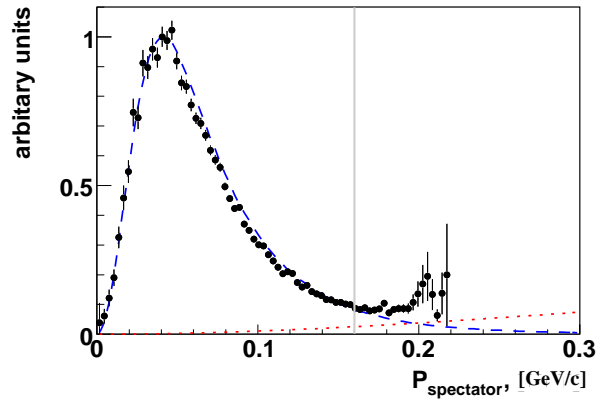}
\caption{\small 
  Distribution of the spectator proton momenta in the $pd \to d\pi^0\pi^0 + 
  p_{spectator}$ reaction. Data are given by solid dots. The dashed line shows
  the expected distribution for the quasifree process based on the CD Bonn 
  potential \cite{mach} deuteron wavefunction. For comparison the dotted line
  gives the pure phase-space distribution 
  as expected for a coherent reaction process. It
  extends up to momenta of 1.5 GeV/c and peaks around 0.7
  GeV/c. For the data analysis only events with $p_{spectator} <$ 0.16 GeV/c
  (vertical line) have been used.
}
\label{fig1}
\end{figure}


\begin{figure} 
\centering
\includegraphics[width=0.99\columnwidth]{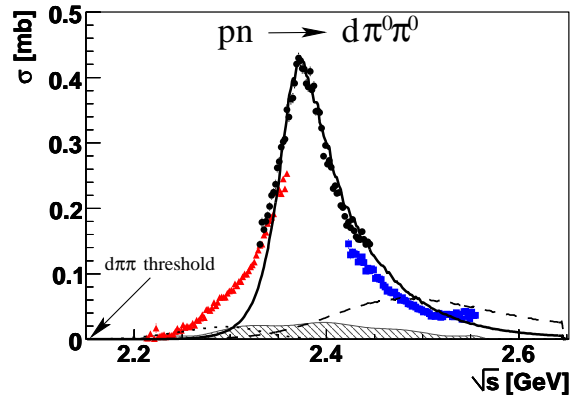}
\caption{\small Total cross sections obtained from this experiment on $pd \to
  d\pi^0\pi^0 +  
  p_{spectator}$ for the beam energies $T_p$ = 1.0 GeV (triangles), 1.2 GeV
  (dots) and 1.4 GeV (squares) normalized independently.
Shown are the total cross section data after acceptance, efficiency and Fermi
  motion corrections.   
  The hatched area indicates systematic uncertainties.
  The drawn lines represent the expected 
  cross sections for the Roper excitation process (dotted) and the $t$-channel
  $\Delta\Delta$ contribution (dashed) as well as a calculation for a
  $s$-channel resonance with m = 2.37 GeV and $\Gamma$ = 68
  MeV (solid).
}
\label{fig1}
\end{figure}

\begin{figure} 
\centering
\includegraphics[width=0.99\columnwidth]{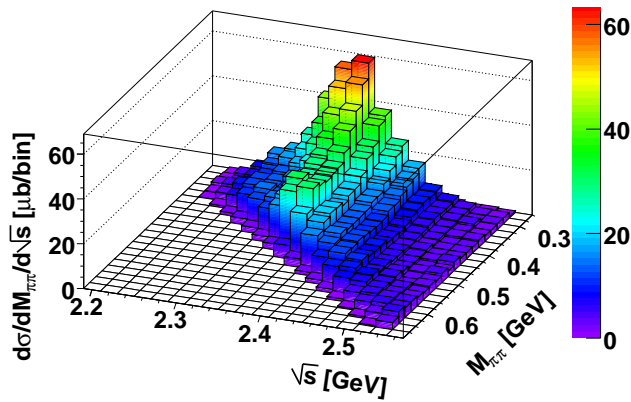}
\caption{\small Energy dependence of the cross section as a function of the
  $\pi^0\pi^0$ invariant mass $M_{\pi^0\pi^0}$ 
  shown by a 3D-plot of the cross section versus $M_{\pi^0\pi^0}$ and $\sqrt{s}$.
}
\label{fig1}
\end{figure}

\begin{figure} 
\centering
\includegraphics[width=0.49\columnwidth]{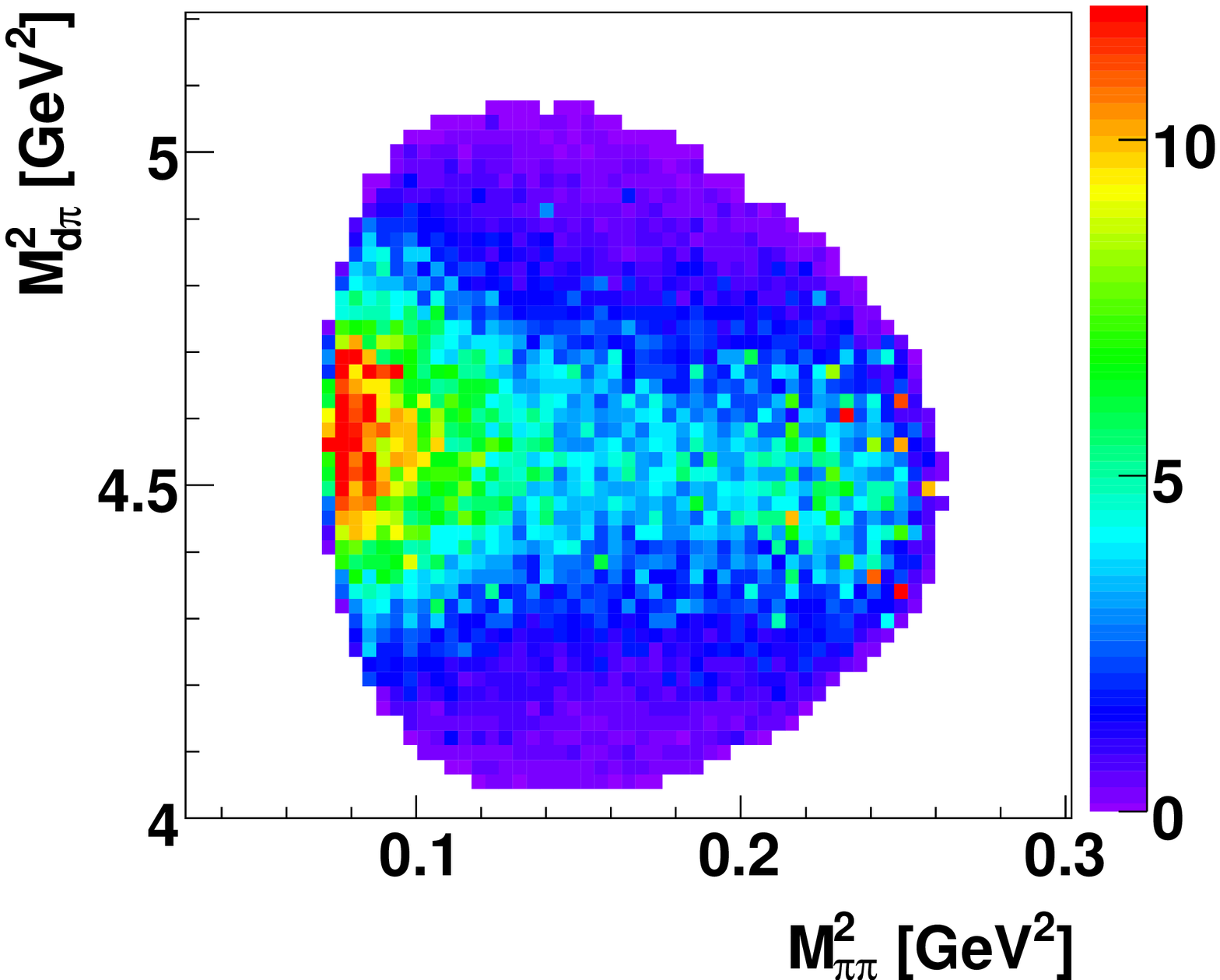}
\includegraphics[width=0.49\columnwidth]{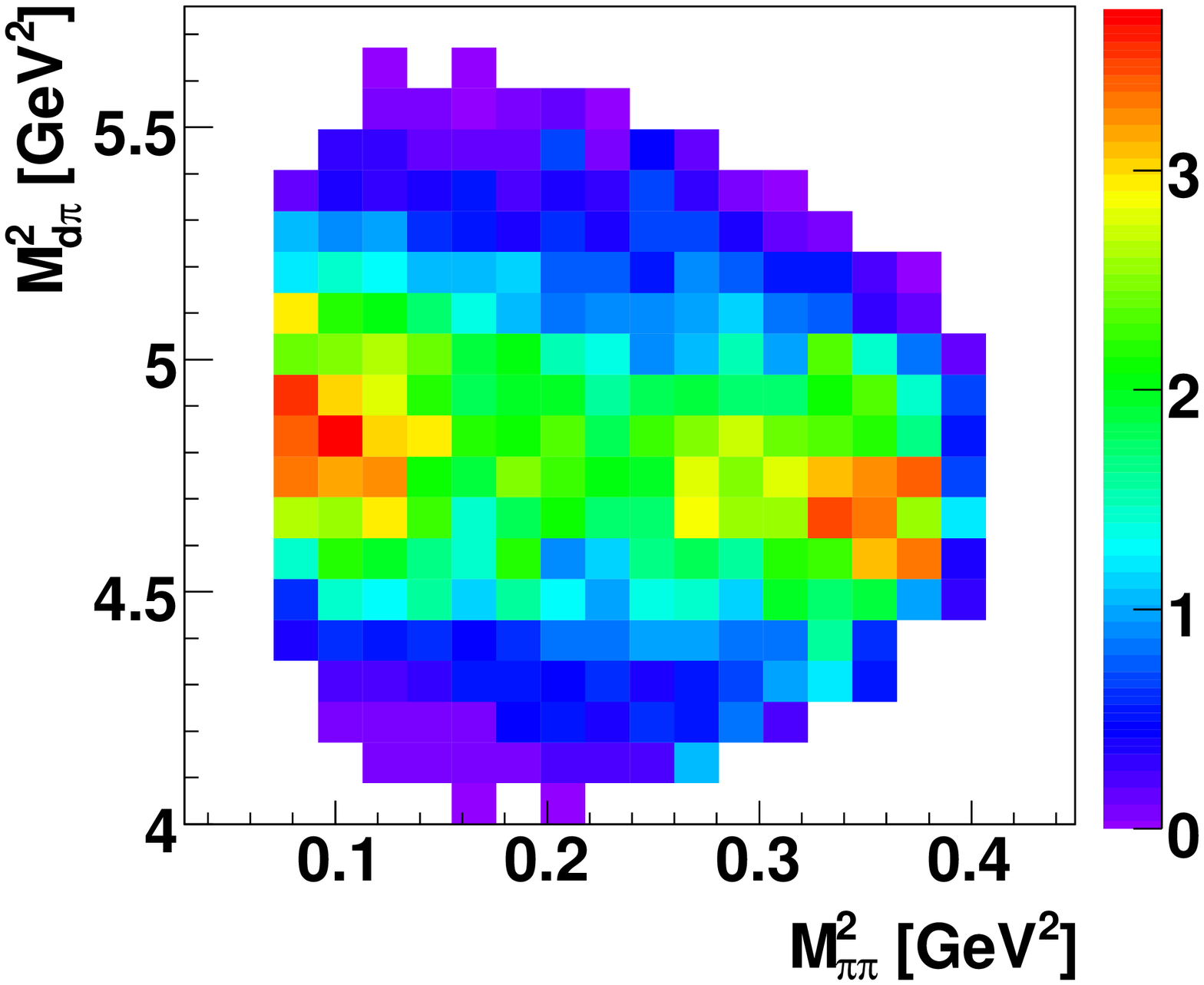}
\includegraphics[width=0.49\columnwidth]{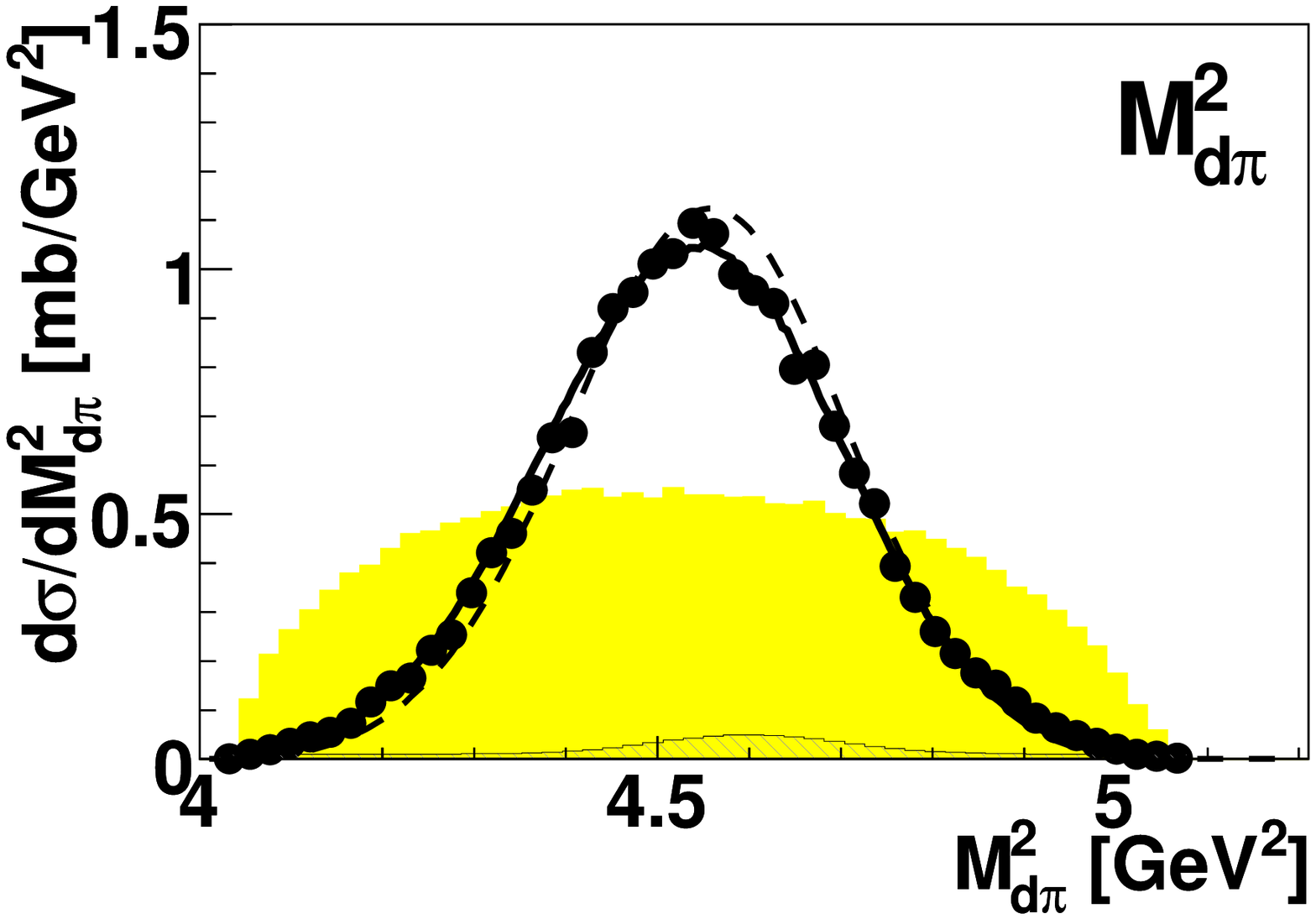}
\includegraphics[width=0.49\columnwidth]{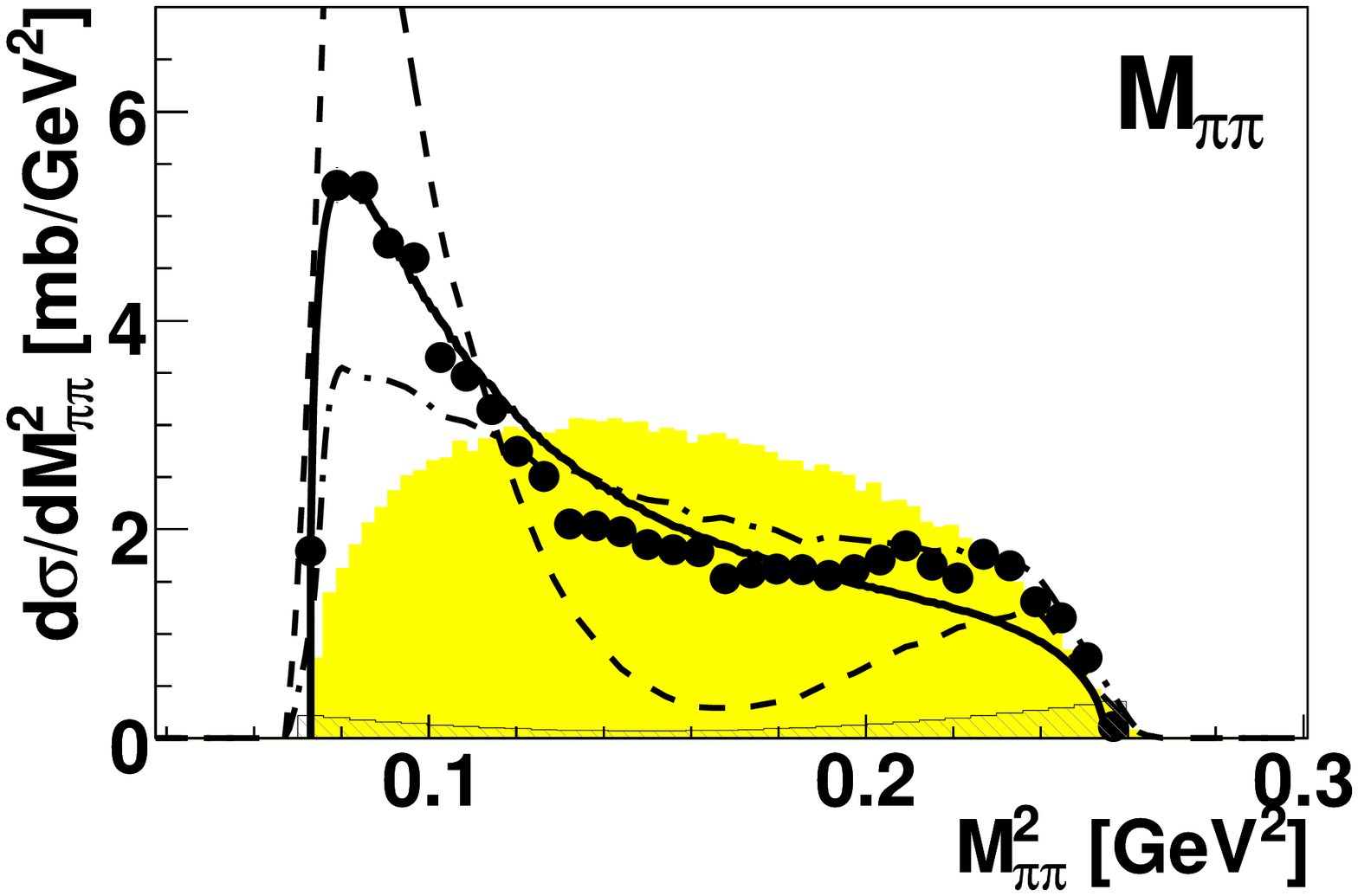}
\caption{\small {\bf Top:}
  Dalitz plots of $M_{d\pi^0}^2$ versus $M_{\pi^0\pi^0}^2$ at $\sqrt{s}$ = 2.38
  GeV (peak cross section) ({\bf left}) and at $\sqrt{s}$ = 2.5 GeV ({\bf
    right}). {\bf Bottom:} Dalitz plot projections $M_{d\pi^0}^2$ ({\bf
    left}) and  $M_{\pi^0\pi^0}^2$ ({\bf right}) axes at  $\sqrt{s}$ = 2.38
  GeV. The curves denote calculations for a $s$-channel resonance
decaying into $\Delta\Delta$ with
  $J^P = 3^+$ with (solid) and without (dash-dotted) form factor as well as
  for $J^P = 1^+$ (dashed). Hatched and
  shaded  areas represent systematic uncertainties and phase-space
  distributions, respectively.  
}
\label{fig:spectra}
\end{figure}

The data obtained with WASA-at-COSY  are in good agreement with those
obtained previously \cite{MB} 
\footnote{In Ref. \cite{MB} a pseudo-Jackson 
  frame was used for  $\theta_\pi^{\pi\pi}$.} at CELSIUS, however, of much
better statistics and precision. Results are shown in Figs. 1 - 5. In the
overlap regions of the 3 measured energy ranges the data agree within their
uncertainties, particularly in absolute magnitude within the normalization
errors. 
Fig. 1 exhibits the momentum distribution obtained for the spectator
proton. It is found to be consistent with that for a
quasifree process. In order to minimize contributions from possible
non-quasifree processes we only use data with $p_{spectator} <$ 0.16 GeV/c.


Fig. 2 shows the energy dependence of the total cross section. 
It exhibits a
very pronounced Lorentzian shaped energy distribution reminiscent of a 
resonance. The width of this structure is four times smaller
than that of a conventional $\Delta\Delta$ excitation process with a
width of about $2\Gamma_\Delta$. Also, the peak
cross section is about 80 MeV below the nominal mass of $2 m_{\Delta}$. The
cross section for the $t$-channel $\Delta\Delta$ process as given in Fig. 2 by
the dotted line is derived by isospin relations from the known $pp \to
d\pi^+\pi^0$ cross section \cite{fk} assuming the absence
of initial state interactions. Based on Ref.~\cite{TS}  the only other major
contribution to this reaction channel should be the Roper excitation process 
(dashed line based on Ref. \cite{alv}
with updated $N^* \to\Delta\pi$ branching ratio
  \cite{skor,bonn}), although there might be other,
non--resonant contributions of relevance~\cite{liu}. 
It seems that
conventional processes contributing to $pn \to d \pi^0\pi^0$ not only are much
smaller in magnitude, but also at variance with the energy dependence of the
data.

\begin{figure} 
\begin{center}
\includegraphics[width=0.98\columnwidth] {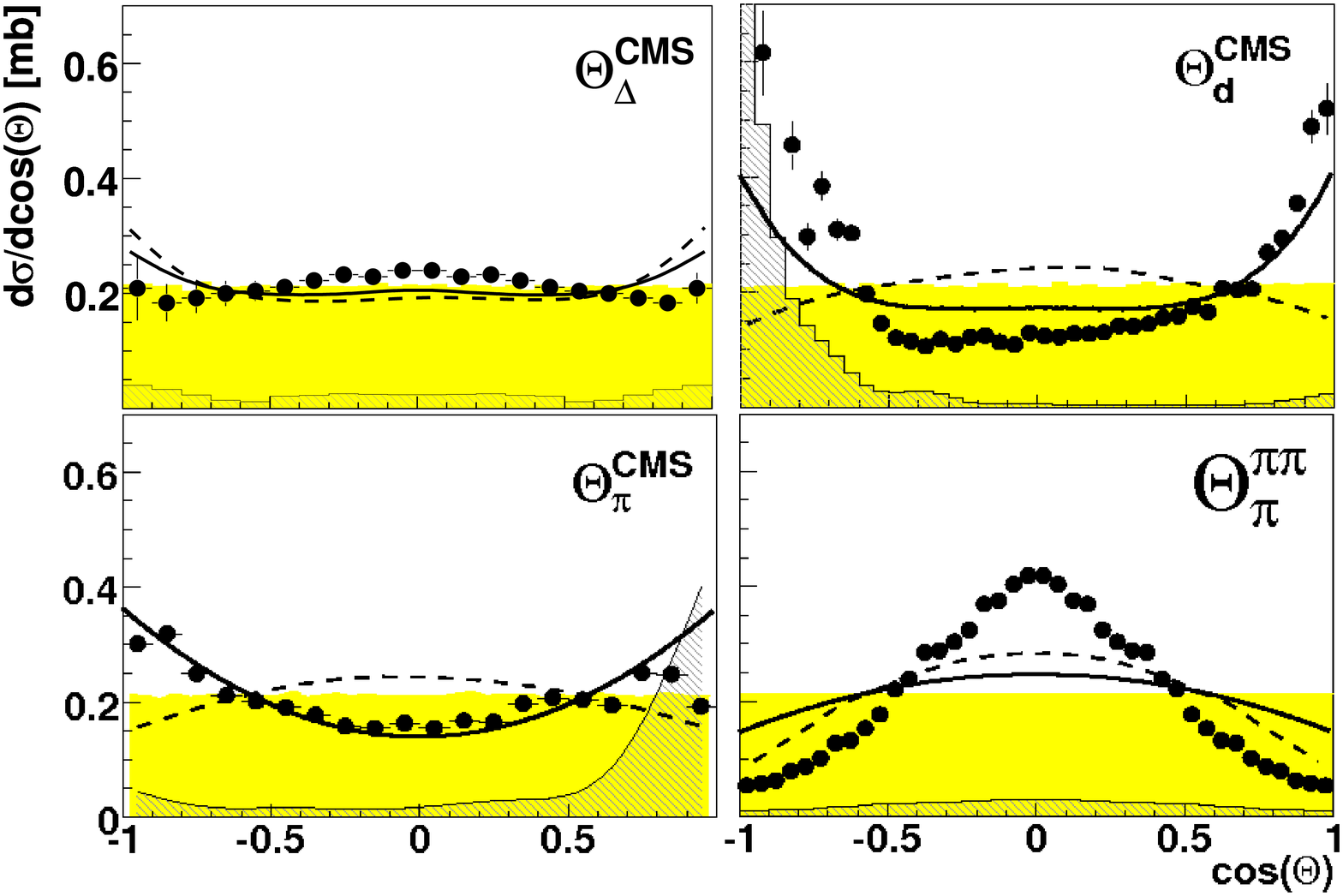}
\caption{\small Same as Fig. 4, but for angular distributions at the peak
  cross section ($\sqrt{s}$ = 2.38 GeV). 
  {\bf Top}: $\Delta$ ({\bf left}) and deuteron ({\bf right}) cms
  distributions, {\bf bottom}: pion distribution in cms ({\bf left}) and Jackson
  frame ({\bf right}).
}
\end{center}
\end{figure}

Fig. 3 displays the cross section in dependence of the $\pi\pi$-invariant mass
$M_{\pi^0\pi^0}$ and the center-of-mass energy $\sqrt{s}$. A 
pronounced low-mass enhancement is observed in the $M_{\pi^0\pi^0}$
distribution -- the ABC effect -- but only at energies within the 
resonance-structure ("ABC region") in the total cross section. Outside this
structure the $M_{\pi^0\pi^0}$ distribution gets rather flat. 

Fig. 4 shows the Dalitz plots of the invariant mass
squared  $M_{d\pi^0}^2$ versus $M_{\pi^0\pi^0}^2$ for two energies: 
at the peak cross section ($\sqrt s$ = 2.38 GeV) and above the ABC
region ($\sqrt s$ = 2.50 GeV). The Dalitz plots exhibit an
enhancement in horizontal direction, in the region of the $\Delta$
excitation, as it prominently shows up in the $M_{d\pi^0}^2$ projection in
Fig. 4, bottom. This feature is consistent with the excitation of a
$\Delta\Delta$ system in the intermediate state --- from now on we assume this
configuration to be realized.
Above the ABC region the Dalitz plot displays only gentle maxima both
at low and high $\pi\pi$  masses, as predicted by the conventional
$t$-channel $\Delta\Delta$ process \cite{ris}. In contrast,
at the peak cross section we see a large enhancement at the
low-mass kinematic limit of $M_{\pi^0\pi^0}^2$. 
 

Angular distributions allow to deduce the total angular momentum of the
reaction. Within the ABC region the angular distributions stay very similar in
shape to those at the peak  cross section. 
Since the deuteron is a loosely bound state the relative momentum of the
two nucleons after pion emission must be small and may be neglected relative
to the pion momenta. Therefore, from the pion and deuteron momenta the 
$\Delta$ momenta can be reconstructed.
Deuteron and $\Delta$ angular distributions in the center-of-mass system (cms)
are displayed in Fig. 5. The very 
low-energetic deuterons going backward in the cms are not completely covered
in our measurements, and thus the systematic uncertainties indicated by the
hatched area in Fig. 5 get very large in this region. Within
uncertainties the angular distributions are symmetric around 90$^{\circ}$ as
demanded for an isospin conserving reaction in a $NN$ system with definite
initial isospin. 
The $\Delta$ angular distribution is essentially isotropic, consistent with
an $s$--wave in the intermediate $\Delta\Delta$ system. This is not unexpected,
since we are largely below the nominal threshold of $2 m_{\Delta}$. 
The deuteron and pion angular distributions are strongly anisotropic.  

In order to proceed it becomes necessary to properly implement the possible
reaction dynamics. We performed a microscopic calculation for an assumed 
resonance state in a given partial wave to decay via $\Delta\Delta$
in an $s$--wave. After the $\Delta$ decays the nucleons merge
to form a deuteron. This model allows us to extract the partial
wave content of the assumed resonance state.

The final state requires the total isospin to be zero. For an $isoscalar$
$\Delta\Delta$ system in  
the intermediate state in a relative $s$-wave, antisymmetrization requires
$J^P = 1^+$ or $3^+$. Hence we confront the data with calculations for
an $s$-channel resonance process $pn \to R \to \Delta\Delta \to d\pi^0\pi^0$
for both spin assignments. 
With the ingredients described up to here, the $M_{\pi\pi}$ distribution
is not yet described well (dash-dotted line in Fig. 4).
This discrepancy
can be improved by including a $\Delta\Delta$ vertex function.
Parameterizing this as a monopole form factor calls for a cut--off scale as
small as $0.15$ GeV/$c^2$, close to the mass of the pion. 
The corresponding fit is shown by solid and dashed lines in Figs. 4 and 5 for
$J^P = 3^+$ and $1^+$, respectively. 
Note, if the relative momenta of the nucleons in the deuteron
were negligible, the argument of that vertex function would be
just the pion relative momentum. This explains the high sensitivity
of the $\pi\pi$ spectrum to this quantity.
In addition to the $M_{\pi\pi}$ distribution the
observed angular distributions for deuterons and
pions in the cms clearly prefer $J$=3 in the ABC region. The
inclusion of background terms, especially the conventional $t$-channel
$\Delta\Delta$ process, further improves the description of the data. 

Our data establish the correlation of the narrow
resonance-like energy dependence with the long-standing puzzle of the ABC
effect.
But so far no conventional process has been identified, which can explain
this phenomenon. In this situation one might 
be tempted to assign the signal to an unconventional $s$-channel resonance
with $I(J^P) = 0(3^+)$, m 
$\approx$ 2.37 GeV and $\Gamma \approx$  70 MeV, as already tentatively
proposed in 
Ref. \cite{MB}. Note that such a resonance has been postulated by various
quark-model calculations, see e.g.  Ref.~\cite{ping}.  With this ansatz
together with the above mentioned vertex function we obtain a good description
of the data (solid lines in Figs. 2 - 5) both in their energy dependence and
in their differential behavior. 
However, to firmly establish
the existence of such a 
resonant system further studies are called
for to provide a microscopic understanding of the unusually small
parameter in the vertex function.
In addition, this resonance
should be observable also in $pn$ elastic scattering. Although its 
effect there is estimated to be only a few percent of the total elastic
cross section, it should be very prominent in the $J^P=3^+$ partial waves.
The present $pn$ data base is very scarce in the
relevant energy range. Thus, precise, polarized measurements in
this channel are urgently called for, in order to allow for a
partial wave analysis.


We acknowledge valuable discussions with L. Alvarez-Ruso, A. Kudryavtsev, 
E. Oset,
A. Sibirtsev,  F. Wang and C. Wil\-kin on this issue. 
This work has been supported by BMBF(06TU9193), Forschungszentrum J\"ulich
(COSY-FFE), 
DFG (683)
and the Foundation for Polish Science (MPD) and EU (Regional Development Fund).

\end{document}